\documentclass[aps,prd,twocolumn,showpacs,superscriptaddress,nofootinbib,preprintnumbers]{revtex4-1}

\usepackage{style}
\usepackage{comment}

\renewcommand{\c}{A}
\newcommand{\kap}{B}
\renewcommand{\S}{\mathcal{S}}

\newcommand{\eps}{\varepsilon}
\newcommand{\Ima}{\mathrm{Im}}

\newcommand{\ybar}{\bar{y}}
\newcommand{\CP}{{\it CP\,}}

\begin{document}

\preprint{ FERMILAB-PUB-26-0289-T, MIT-CTP/6072}

\newcommand{\mytitle}{Baryogenesis via the CKM Matrix with Minimal Flavor Violation}
\title{\mytitle}

\author{Innes Bigaran}
 \affiliation{Department of Physics \& Astronomy, Northwestern University, 2145 Sheridan Road, Evanston, IL 60208, USA}
\affiliation{Theoretical Physics Department, Fermilab, P.O. Box 500, Batavia, IL 60510, USA}
\affiliation{Center for Neutrino Physics, Virginia Polytechnic Institute and State University,
Blacksburg, VA 24061, USA}

\author{Gordan Krnjaic}
\affiliation{Theoretical Physics Department, Fermilab, P.O. Box 500, Batavia, IL 60510, USA}
\affiliation{Department of Astronomy and Astrophysics, University of Chicago, Chicago, IL 60637}
\affiliation{Kavli Institute for Cosmological Physics, University of Chicago, Chicago, IL 60637}

\author{Kevin Langhoff}
\affiliation{Center for Theoretical Physics, Massachusetts Institute of Technology, Cambridge, MA 02139, USA}

\author{Huangyu Xiao}
\affiliation{Theoretical Physics Department, Fermilab, P.O. Box 500, Batavia, IL 60510, USA}
\affiliation{Kavli Institute for Cosmological Physics, University of Chicago, Chicago, IL 60637}
\affiliation{
Physics Department, Boston University, Boston, MA 02215, USA
}
\affiliation{Department of Physics, Harvard University, Cambridge, MA 02138, USA
}

\date{\today}

\begin{abstract}
It is often claimed Standard Model \CP violation is insufficient for baryogenesis. We present a counterexample using minimal flavor violation (MFV) in which all {\it CP}-violating effects arise from the Cabibbo-Kobayashi-Maskawa (CKM) matrix. Our scenario involves a leptoquark field with MFV-preserving interactions whose decays to Standard Model particles yield the observed baryon asymmetry in the early universe. Unlike previous efforts to realize baryogenesis through the \CP violation of the CKM matrix, our scenario does not require any time-variation of model parameters.
\end{abstract}

\maketitle

\section{Introduction}

The origin of the baryon asymmetry is among the major unsolved problems in fundamental physics. From cosmic microwave background and big bang nucleosynthesis data, the observed baryon yield is \cite{Planck:2018vyg} 
\be
\label{eq:eta}
Y^{\rm obs}_B = \frac{n_B-n_{\bar B}}{s} = (8.7 \pm 0.1) \times 10^{-11}~~,
\ee
where $n_B, n_{\bar B}$, and $s$ are baryon, antibaryon, and entropy densities, respectively. If inflation set initial conditions for the hot big bang, baryon number would be zero at reheating.\footnote{Baryon asymmetry may survive inflation given exponential fine-tuning~\cite{Krnjaic:2016ycc}, but is strongly constrained by isocurvature \cite{Murai:2023ntj}.} However, a baryon asymmetry can dynamically arise from a symmetric initial state if particle reactions are out of equilibrium, exhibit $C$ and \CP violation, and violate baryon number, as first noted by Sakharov~\cite{Sakharov:1967dj}. 

The Standard Model (SM) contains all the ingredients required to satisfy Sakharov's conditions: Hubble expansion provides departure from equilibrium, baryon number is violated by electroweak sphalerons, and $C$ and \CP violation have been experimentally observed in meson systems through mixing and decay processes. All quark-sector {\it CP}-violating observables are proportional to the basis-independent Jarlskog  invariant~\cite{ParticleDataGroup:2024cfk}
\be
J = {\rm Im}\!\left(V_{ud} V_{cs} V_{us}^* V_{cd}^*\right) = 
\left(3.12_{-0.12}^{+0.13}\right) \times 10^{-5} ,
\ee
where $V$ is the Cabibbo-Kobayashi-Maskawa (CKM) matrix. However, in the limit where any pair of equal-charge quarks is degenerate in mass, the SM CKM phase becomes unphysical and the aforementioned {\it CP}-violating effects disappear~\cite{Jarlskog:1985ht,Jarlskog:1986mm}. Thus, any baryon asymmetry generated from SM CP violation naively scales as\footnote{This depends on the renormalization-group (RG) scale. Below, we use ratios of Yukawas evaluated at $200$~GeV \cite{Alam:2022cdv} to reduce renormalization group-scale ambiguities.}
\begin{align}
\label{eq:YB_EWmain}
Y_B \sim  J\prod_{i<j}  (y_{u_i}^2-y_{u_j}^2) \prod_{k<l}(y_{d_k}^2-y_{d_l}^2) \lesssim 10^{-20},~~~~
\end{align}
which is many orders of magnitude smaller than Eq.~\eqref{eq:eta}. This discrepancy underlies the intuition that SM \CP-violation is insufficient for baryogenesis~\cite{Gavela:1993ts, Gavela:1994ds,Gavela:1994dt,Huet:1994jb}. 

However, the suppression in Eq.~\eqref{eq:YB_EWmain} can be circumvented under the right conditions. One possibility is that the SM Yukawa couplings had markedly larger values in the early universe ~\cite{Berkooz:2004kx,Perez:2005yx}; however, such models destabilize the Higgs potential~\cite{Braconi:2018gxo} and therefore require additional new physics. Moreover, the scaling of \CP \, violation with differences in Yukawa couplings only arises when the universe is approximately flavor-blind\footnote{This is tied to the breaking of flavor symmetry at lower energies where the SM quarks are massive, see App.~\ref{app:inv} for a more detailed discussion.} ({\it i.e.} $T\gg m_t$), so lower-energy processes can break this proportionality. This feature is exploited in models of Mesogenesis, where the required \CP \, violation arises from enhanced $B$-meson oscillations at low temperatures~\cite{Elor:2018twp,Nelson:2019fln,Elor:2020tkc,Davoudiasl:2026qba}.  However, reproducing the observed baryon asymmetry within this framework requires new particles which are light in the early universe, but become heavy at late times in order to evade laboratory constraints~\cite{Elor:2018twp,Nelson:2019fln,Elor:2020tkc,Davoudiasl:2026qba}.

In this {\it Letter}, we introduce a novel model that realizes baryogenesis using only the \CP \, violation of the SM, without the need for time-varying parameters. Our setup extends the SM with a scalar leptoquark whose baryon-violating interactions are governed by minimal flavor violation (MFV) \cite{Buras:2000dm, DAmbrosio:2002vsn}, and whose decays can yield the observed baryon asymmetry without any additional sources of \CP \,violation beyond the CKM phase.\footnote{Our MFV-based approach to baryogenesis is analogous to models of leptogenesis under minimal lepton flavor violation~\cite{Cirigliano:2006nu,Branco:2006hz}.}

\section{Model Overview}
\label{sec:model}
We introduce a scalar leptoquark $\cal S$, which carries one of the three SM gauge representations that allow baryon and lepton  violating interactions at tree level.\footnote{These SM gauge ($SU(3)_c \otimes SU(2)_L \otimes U(1)_Y$) representations are $({\mathbf 3},1)_{-1/3}, ({\bf 3},1)_{-4/3}$, and  $({\bf 3},{\bf 3})_{-1/3}$; see Ref.~\cite{Dorsner:2016wpm} for a discussion. Note that a fourth baryon-number violating scalar exists in the presence of right-handed neutrinos.} Additionally, we charge $\cal S$ under the global quark-flavor group 
\begin{align} 
{\cal G}_{F}= SU(3)_Q \otimes SU(3)_{\bar{u}} \otimes SU(3)_{\bar{d}}~,
\end{align} 
which is explicitly broken by SM Yukawa interactions
\be
-\mathcal{L}_Y \supset Y_u^{Aa} H Q_A \bar u_a +Y_d^{Ai} H^\dagger Q_A \bar d_i + h.c.~,
\ee
where fundamental representations of $SU(3)_Q$ are labeled with $(A,\,B,\,C,...)$, $SU(3)_{\bar{u}}$ with $(a,\,b,\,c,...)$, and $SU(3)_{\bar{d}}$ with $(i,\,j,\,k,...)$. The SM $Q$, $\bar u$, and $\bar d$ fields transform as fundamentals under their respective $SU(3)$ groups, so flavor symmetry is formally restored by promoting the Yukawa couplings to spurions, which transform as $Y_u=(\bar{\bm{3}}_Q,\bar{\bm{3}}_{\bar{u}})$ and $Y_d =(\bar{\bm{3}}_Q,\bar{\bm{3}}_{\bar{d}})$ under ${\cal G}_F$.

Flavor breaking interactions are technically natural and are faithfully transmitted from the UV to low energies. If the ultraviolet~(UV) completion of the SM only has a single source of flavor breaking, then Yukawa spurions fully source all flavor violation at lower scales. The framework of \emph{minimal flavor violation} (MFV) elevates this concept to a postulate: even in the presence of new physics, $Y_{u,d}$ are assumed to be the only sources of flavor violation and all flavor structures must be built from functions of $Y_{u,d}$ \cite{Buras:2000dm, DAmbrosio:2002vsn}. 

We assume that all $\cal S$ interactions preserve MFV, so the Lagrangian contains terms of the form 
\begin{align}
    \mathcal{L}_{\S} = f(Y_u,Y_d) \,  \S \, \mathcal{\hat O}_{\rm SM}~,~
\end{align}
where $\mathcal{\hat O}_{\rm SM}$ is a SM operator, $f(Y_u,Y_d)$ is a function of quark Yukawa spurions (and their conjugates), and all gauge and flavor indices have been suppressed. In order to preserve MFV, $\mathcal{L}_{\S}$ is required to be a ${\cal G}_F$ singlet such that the charge of $\S$ under ${\cal G}_F$ determines the compensating form of $f(Y_u,Y_d)$ for each choice of $\hat {\cal O}_{\rm SM}$.

To demonstrate essential features of this scenario, consider a proof-of-principle in which $\S$ transforms as  $(\bar{\bm{3}},\bm{1})_{1/3}$ under the SM gauge group and as a $\mathbf{3}_{\bar d}$ under   ${\cal G}_F$. For this choice of charge assignments, the renormalizable and gauge-invariant $\S$-SM interactions are 
\begin{align}
\label{eq:lag_X}
{\cal L}_{\cal S} =  {\c}_{i j a}  \S^i  \bar d^j \bar{u}^a  +  \kap^i_{a} \S^\dagger_{i} \bar u^a \bar e  + h.c.~~,
\end{align} 
where $A_{ija}$ and $B_a^i$ are matrices in flavor space and 
we do not impose lepton flavor symmetries; $\bar e$ is a ${\cal G}_F$ singlet and could be any charged lepton flavor.\footnote{An extension of our analysis would be to consider lepton flavor more carefully, motivating the elimination or significant suppression of certain decay channels. Moreover, one could consider grand unified theories where SM fermions are embedded into representations that relate leptons and quarks.}
To be concrete, we can choose 
\begin{align} 
    \c_{ija} &=  \alpha \; \epsilon^{ABC}(Y_d)_{Ai} (Y_d)_{Bj} (Y_u)_{Ca}\label{eq:c} \\
    \kap^i_{a}~ &= \beta \;\epsilon_{ABC} \epsilon_{abc}(Y_d^\dagger)^{iA} (Y_u^\dagger)^{bB}(Y_u^\dagger)^{cC} \, , \label{eq:kap}
\end{align}
where $\alpha$ and $\beta$ are real-valued, as we have freedom to rotate SM fields to absorb non-physical phases. Color indices are omitted for brevity from eq.~\eqref{eq:lag_X} but included for subsequent calculations.\footnote{ When restored, these color indices are denoted Greek letters and read $  {\c}_{i j a}  \S^i  \bar d^j \bar{u}^a\equiv  \epsilon_{\alpha\beta\gamma}{\c}_{i j a}  \S^{i\alpha}  \bar d^{j\beta} \bar{u}^{a\gamma}$   and   $\kap^i_{a} \S^\dagger_{i} \bar u^a \bar e\equiv \delta^\alpha_{\beta}\kap^i_{a} \S^\dagger_{i, \alpha} \bar u^{a\beta} \bar e$. 
} Note that unitarity places upper bounds on $\alpha$ and $\beta$:
\begin{align}
    |\alpha_{\rm max}| \approx \frac{2\sqrt{\pi}}{y_s\, y_b\, y_t\, |V_{td}|}~~,~~
    |\beta_{\rm max}| \approx \frac{\sqrt\pi}{y_s\, y_c\, y_t\, |V_{us}|},
\end{align}
 which are obtained by imposing ${\rm max}(A_{ija}) = {\rm max}(B^i_a) = \sqrt{4\pi}$ and, for convenience, we also define
 \begin{align}
 \label{eq:ralphabeta}
 r_\alpha \equiv |\alpha|/\alpha_{\rm max}~~,~~r_\beta \equiv|\beta|/\beta_{\rm max},
 \end{align}
 which saturate to $r_{\alpha,\beta} = 1$ at the unitarity limit.

In  the $Y_{u,d} \to 0$ limit,  ${\cal G}_F$ symmetry requires all ${\cal S}_i$ components to have degenerate masses. However, once flavor is broken,
these scalars generically acquire  mass terms of the form
\begin{align}
\label{eq:masses}
-{\cal L}_M =  M^2(\delta^i_{\, j}
+\Delta^i_{\, j}) \,\S^\dagger_i \S^j ~,
\end{align}
where ${\Delta^i}_{ \, j}$ is an $\bm{8}_{\bar{d}}$ spurion, which we truncate at quadratic order in SM Yukawas
\begin{align}
    {\Delta^i}_{\!j}= \gamma \, (Y_d^\dagger Y_d)^i_{\ j}\,,\quad \gamma \in \mathbb{R}\,.
\end{align}
Throughout this paper, we 
work in the basis where this term is diagonal, so the squared ${\cal S}_i$ mass becomes
\be
\label{eq:mass-eigen}
m_i^2 = M^2(1+\gamma\, y_{d_i}^2),
\ee
and any renormalization group-scale ambiguity can be absorbed into the definition of $\gamma$. Since the mass terms preserve the flavor symmetry for $\gamma \to 0$, we generically expect $\gamma \ll 1$ on technical naturalness grounds. Note also that the non-degeneracy of the $m_i$ allows leptoquark-induced $\it CP$ \,violation to be appreciable at high temperatures $(T \gg m_t)$ even when SM-only interactions are approximately flavor blind and their {\it CP}-violation is sharply suppressed (see App.~\ref{app:inv}).

\section{Baryon-Violating Decays}
\label{sec:bv-decays}

\begin{figure*}[t!]
\includegraphics{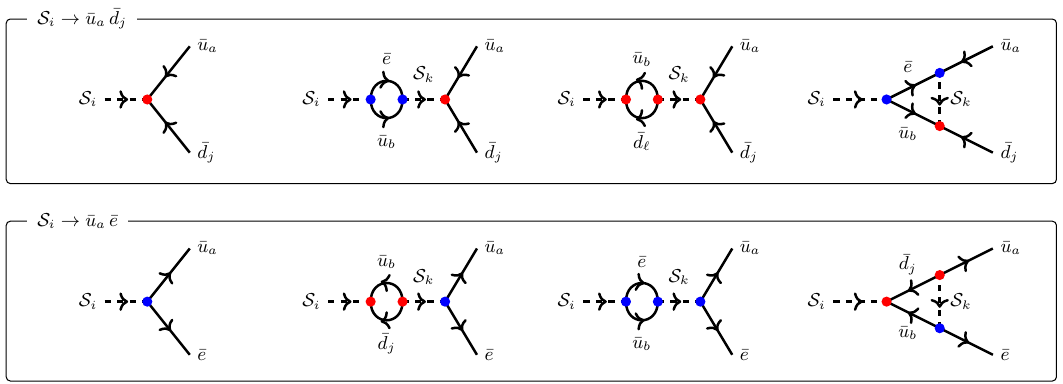}
    \caption{Leading Feynman diagrams for $\S_i \to \bar{u}_a^\dagger ~\bar{d}_j^\dagger$ (top) and  $\S_i \to \bar{u}_a \bar{e} $  (bottom) decay channels, where the $A_{ija}$ and $B^i_a$ vertices from Eq.~\eqref{eq:lag_X} are shown as \textbf{\textcolor{red}{red}} and \textbf{\textcolor{blue}{blue}} dots, respectively.}
    \label{fig:BV_Decay_Diagrams}
\end{figure*}

To generate a net baryon asymmetry, $\mathcal{S}$ decays to baryonic final states must differ from their {\it CP}-conjugates. The net baryon yield per ${\cal S}_i$ decay is
\be
    \epsilon_i  = \sum_{F}B_F \;\frac{
    \Gamma( \mathcal{S}_i \rightarrow F ) - \Gamma( \mathcal{S}^\dagger_i \rightarrow \bar F ) }{\Gamma_i }
    \label{eq:epsilon_i},~~~~~
\ee
where $\Gamma_{i}$ is the total ${\cal S}_i$ width and the sum is over decay channels $F$ with baryon number $B_F$; a baryon asymmetry arises from the interference between the tree and loop-level diagrams shown in Fig.~\ref{fig:BV_Decay_Diagrams}. In the massless fermion limit, the amplitude for ${\cal S}_i$ decays to final state $F$ is 
\be 
    \mathcal{M}( \mathcal{S}_i \!\rightarrow \! F ) =  C^{i}_{T,F}\, \mathcal{A}_T^{i} + \!\sum_k\left(C_{P,F}^{ik}\, \mathcal{A}_P^{ik} + C^{ik}_{V,F}\, \mathcal{A}_V^{ik}\right)\!,~~~~~
\ee
where $k$ is the flavor of $\mathcal{S}$ appearing in the loop, $T$ denotes a tree-level process, and $P$/$V$ denote 1-loop diagrams correcting the propagator and vertex, respectively. The $C$ coefficients collect the Yukawa couplings and color factors, while the $\mathcal{A}$ factors carry the kinematics.

In App.~\ref{app:Calculation of Baryon Asymmetry} we show the leading contribution to the asymmetry is
\be
\label{eq:asymm-factor}
    \epsilon_i= -\frac{4\sum_{F}B_F \sum_{L}\sum_k \operatorname{Im}(C^i_{T,F}\, C^{ik*}_{L,F}) \operatorname{Im}(\mathcal{A}^i_T \mathcal{A}^{ik *}_L)}{\sum_{F}|C^i_{T,F}|^2|\mathcal{A}^i_T |^2},\> ~~~~~
\ee
where $L \in \{P,V\}$. In the down-aligned basis (see App.~\ref{app:Calculation of Baryon Asymmetry} for details), the terms in the denominator are
\begin{align}
\label{eq:Cud}
    &|C^i_{T,ud}|^2 =2 |\alpha|^2 \sum_{m\neq i}\sum_a\bar y_{d_m}^2y_{u_a}^2 |V_{am}|^2\\
    &|C^i_{T,ue}|^2 = 4|\beta|^2y_{d_i}^2\sum_{a} \bar y_{u_a}^2 |V_{ai}|^2~,
    \label{eq:Cue}
\end{align}
where, for future convenience, we have defined
\begin{align}
\ybar_{d_i} \equiv \prod_{j\neq i}y_{d_j}\,,\qquad
\ybar_{u_a} \equiv \prod_{b\neq a}y_{u_b}\,,
\end{align}
so the total decay width for ${\cal S}_i$ is
\begin{align}
    \Gamma_i
    &= 
    \frac{1}{16\pi m_i}\sum_{F}|C^i_{T,F}|^2|\mathcal{A}^i_T |^2\notag \\ 
    &\approx 
    \frac{m_i}{16 \pi }\left(|C^i_{T,ud}|^2 + |C^i_{T,ue}|^2\right).~~~~~
\end{align}
where the color multiplicity of each channel is already contained in $|C^i_{T,F}|^2$, Eqs.~\eqref{eq:Cud} and \eqref{eq:Cue}. Note that the numerator in Eq.~\eqref{eq:asymm-factor} vanishes for $T$-$V$ interference (see App.~\ref{app:Calculation of Baryon Asymmetry}),  while $T$-$P$ interference gives 
\begin{align} \label{eq:C_TP}
    \operatorname{Im}(C^i_{T,ud}\, C^{ik*}_{P,ud}) &= - \operatorname{Im}(C^i_{T,ue}\, C^{ik*}_{P,ue}) \\
    &=-8|\alpha|^2|\beta|^2\,\varepsilon_{ikr}\,y_d^2\, y_s^2\, y_b^2\,  y_t^4 \,y_c^2 \, J \, ,
\end{align}
where $r$ is the single index in $ \{1,2,3\} \setminus \{i,k\}$. The 1-loop kinematic term is
\begin{align}
     \operatorname{Im}(\mathcal{A}^i_T \mathcal{A}^{ik *}_L) = \frac{|\mathcal{A}^i_T|^2}{16 \pi } f_L^{ik} ,
\end{align}
where  $f_L$ is a kinematic function for each loop topology. As only the $L = P$ term contributes, 
\begin{align}
\label{eq:fPik}
f_P^{ik} = \frac{ m_i^2(m_k^2-m_i^2) } {(m_k^2-m_i^2)^2 +m_k^2 \Gamma_k^2},
\end{align}
which vanishes in the $\gamma \to 0$ limit where $m_i = m_k$. 

\begin{figure*}[t]
    \centering
    \includegraphics[width=1\linewidth]{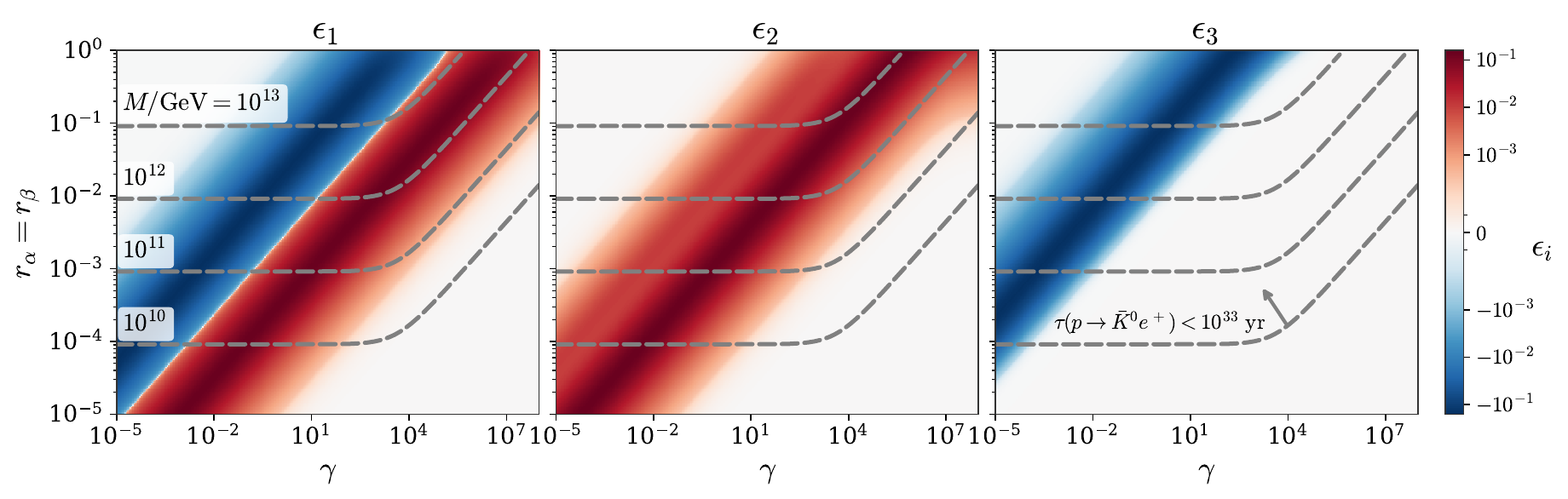}
    \caption{Density plots of each baryon-asymmetry efficiency  $\epsilon_i$  as a function of coupling strength $r_{\alpha,\beta}$ from Eq.~\eqref{eq:ralphabeta} and the mass-splitting parameter $\gamma$ from Eq.~\eqref{eq:mass-eigen}. Note that $r_{\alpha,\beta}$ are coupling strengths normalized to their unitarity maxima, as defined in Eq. \eqref{eq:ralphabeta}. The grey curves show the upper bound on $r_\alpha = r_\beta$ vs. $\gamma$ for several values of $M$ from constraints on $p^+\to \bar{K}^0 e^+$ (Sec.~\ref{sec:Proton_Decay}). Importantly, this proton-decay channel is only kinematically allowed if ${\cal S}_i$ couples to electrons in Eq.~\eqref{eq:lag_X}, which is not required; if the lepton is $\tau$ instead of $e$, the proton-decay bound is significantly relaxed relative to what is shown here.}
    \label{fig:eps_contour}
\end{figure*}

Near resonance, when $(m_k^2-m_i^2)^2 \approx m_k^2 \Gamma_k^2$, one finds $|f_P^{ik}| \to m_i^2/(2m_k\Gamma_k) \simeq 8\pi/{\cal D}_k$, where ${\cal D}_i \equiv |C^i_{T,ud}|^2+|C^i_{T,ue}|^2$. Thus, the maximal asymmetry per ${\cal S}_i$ decay becomes 
\begin{align}
    \epsilon^{\rm max}_{i} \approx \frac{64 \pi^2 r_\alpha^2 r_\beta^2  J}{{\cal D}_i{\cal D}_k} \frac{y_d^2/y_s^2 }{|V_{td}|^2\,|V_{us}|^2}, 
\end{align}
where $k$ is the resonant partner of $i$. Let us consider for algebraic simplicity the condition $r_\alpha = r_\beta$ which causes $r_\alpha$ and $r_\beta$ to cancel between numerator and denominator. As an explicit example, consider $i = 1$ where ${\cal D}_1{\cal D}_k \approx 9\pi^2\, r_\alpha^4$ for $r_\alpha = r_\beta$ and either $k = 2,3$; this gives
\begin{align}
\label{eq:eps1_max}
    \epsilon^{\rm max}_{1} &\approx  
    \frac{64\, J}{9}
    \frac{y_d^2/y_s^2}{|V_{td}|^2\,|V_{us}|^2}
    \approx 0.15\,,
\end{align}
where we used the baryon number assignments
$B_{\bar u^\dagger \bar d^\dagger} - B_{\bar u \bar{e}} = 1$. More generally, in Fig.~\ref{fig:eps_contour} we show a density plot of each $\epsilon_i$ in the $r_\alpha -\gamma$ plane, again assuming $r_\alpha = r_\beta$ for simplicity.

\section{Cosmology}

Thus far, we have shown that a symmetric population of $\S$ and $\S^\dagger$ particles can produce a large average baryon asymmetry per particle decay. However, these decays preserve $B-L$, so the resulting asymmetry is not protected from washout by $B+L$ violating electroweak sphalerons \cite{Kuzmin:1985mm}, which are in equilibrium from $T\sim10^{12}$~GeV until decoupling at $T\sim130$~GeV. Thus, baryogenesis must either occur after sphalerons decouple or generate a nonzero $B-L$ asymmetry at earlier times; here we discuss both possibilities in turn.

\subsection{\bf Low Reheat Temperature}
\label{sec:lowTRH}
As we will see below in Sec.~\ref{sec:Proton_Decay}, the $S_i$ must be quite heavy to avoid proton decay bounds. 
If the universe is reheated below the weak scale to avoid sphaleron washout, then these much heavier $S_i$ must be produced non-thermally and decay out of equilibrium to generate the baryon asymmetry. Here we present a simple scenario which viably realizes this cosmological history.

Consider a population of long-lived scalars $\Phi$, which dominate the early universe and decay via $\Phi \to \S_i \S_i^\dagger$ reactions before big bang nucleosynthesis. Subsequently, the $S_i$ decay promptly and  reheat the universe to a temperature $T_{\rm RH} < 130$~GeV. In the instantaneous decay approximation,  energy conservation yields
\be
\label{eq:TRH}
m_\Phi n_\Phi = \frac{\pi^2 g_{\star}}{30} T_{\rm RH}^4~, ~
\ee 
where $m_\Phi$ and $n_\Phi$ are respectively the $\Phi$ mass and number density, and
$g_\star$ is effective number of relativistic species at $T_{\rm RH}$.

The baryon asymmetry produced from these out-of-equilibrium decays is the sum of contributions from each flavor. 
If $\Phi$ populates $\S_i$ with branching ratio $\text{Br}_i$, then
\begin{align}
    Y_B \approx \sum_i \frac{n_{\mathcal{S}_i} \epsilon_i }{s} \approx \frac{3T_{\rm RH}}{4m_{\Phi}} \sum_i \text{Br}_i \,\epsilon_i\,.~\label{eq:YB_exact}
\end{align}
For $g_{\star }\approx \mathcal{O}(100)$, $2m_\S< m_\Phi$, and $\sum_i \text{Br}_i \,\epsilon_i  \lesssim \mathcal{O}(0.1)$, the maximal asymmetry in this low reheating scenario is
\be
    Y_B^{\rm max} \approx 10^{-10}
    \brac{\epsilon}{0.1}
    \left( \frac{10^{11}{\,\rm GeV}}{m_\S} \right)\left(\frac{T_{\rm RH}}{100~{\rm GeV}}\right).~~~
\ee
Since we require $T_{\rm RH}\lesssim 130 \, {\rm GeV}$ to avoid washout from EW sphalerons, generating the observed asymmetry requires $m_\S \lesssim 10^{11}$~GeV; in Sec.~\ref{sec:Proton_Decay} we show that such masses can satisfy proton decay bounds.

We note that there are other possibilities for realizing the low-reheat scenario. For example, the $S_i$ can be produced through the Hawking evaporation  of primordial black holes \cite{Baumann:2007yr,Hooper:2020otu,Bernal:2022pue}. In this variation, the baryon yield depends on the black hole masses and abundances in the early universe, but exploring this scenario is beyond the scope of this work. 

\subsection{High Reheat Temperature}

\label{sec:highTRH}

Alternatively, consider a cosmological history with $T_{\rm RH} \gtrsim m_i$ such that $\S_i$ are initially in chemical equilibrium with the SM plasma, but freeze out while still relativistic. After freeze out, the $\S$ decay out of equilibrium  {\it before} sphalerons equilibrate at $T \sim 10^{12}$~GeV, and these decays produce a pure $B+L$ asymmetry with $B-L = 0$. However, if additional interactions erase $L$ and leave $B$ untouched before $T \sim 10^{12}$~GeV, the universe acquires a net $B-L$ number, which sphaleron interactions do not erase.

A simple realization of this mechanism extends the model in Eq.~\eqref{eq:lag_X} with right-handed neutrinos, whose Majorana masses violate lepton number. If these particles are in equilibrium  with the SM in between $\S$ decay and sphaleron equilibration, their 
$\Delta L = 2$ interactions efficiently erase lepton number while preserving baryon number \cite{Fukugita:2002hu}. These $\Delta L = 2$ interactions must freeze out before sphalerons equilibrate; otherwise simultaneous $L$ and $(B+L)$ violation wash out both $B$ and $L$ numbers. 

This necessary sequence of events is naturally achieved with viable model parameters. For example, if right-handed neutrino masses are well above $10^{12}$~GeV, these processes decouple before $T=10^{12}$~GeV provided that  ${\rm Tr}(m_\nu^\dagger m_\nu) \lesssim (0.8~{\rm eV})^2$, which is consistent with current bounds~\cite{KATRIN:2024cdt}.
Importantly, the final baryon asymmetry can still depend only on quark-sector \CP violation, provided that 
\begin{enumerate}
    \item The additional $B-L$ asymmetry generated from right-handed neutrino decays  is negligible.
    \item Lepton number washout is efficient ($r_L \ll1$ where $r_L$ is the surviving fraction of deposited $L$ number from $\S$ decays after $\Delta L = 2$ processes take effect).
\end{enumerate}
Under these assumptions, the baryon asymmetry is
\begin{align} \label{eq:YB_high_main}
Y_B \approx \frac{28}{79}\,\kappa\,(1-r_L)\sum_i Y_{\S_i}\,\epsilon_i ~~~ (T_{\rm RH} \gtrsim m_i) 
\end{align}
where 28/79 is the sphaleron conversion of $B-L$ into $B$~\cite{Weinberg:2008zzc} and $\kappa$ is the fraction of $B+L$ converted to $B-L$. Note that the deposited lepton asymmetry after $\S$ decays resides in the electroweak singlet $\bar{e}$, which does not participate in $\Delta L = 2$ processes; solving for the chemical equilibrium obtained using $\Delta L = 2$ and top-Yukawa mediated processes gives (see App.~\ref{app:highTRH_window})
\begin{align}
    \kappa = \frac{2}{3}\left(\frac{{\rm rank}(m_\nu)}{{\rm rank}(m_\nu) + 1}\right)\,,
\end{align}
where ${\rm rank}(m_\nu)$ is the rank of the SM neutrino mass matrix which is $2$ or $3$ depending on whether the lightest neutrino is massless.
Although the $\S_i$ population freezes out before they decay, their yields stay sizable so long as $T_{\rm RH}\gtrsim m_i$. At these temperatures even strong interactions struggle to hold $\S_i$ in equilibrium, so decoupling occurs for $m/T=\mathcal{O}(1)$ rather than the familiar $m/T\simeq 20$ ratio for weak-scale freeze-out \cite{Kolb:1990vq}. The resulting yield is
\begin{align}
    Y_{\S_i} &\sim   10^{-2}\brac{m_i}{2\times 10^{14}\,{\rm GeV}}^p; 
\end{align}
$p=1$ for $m_i \lesssim  2\times 10^{14}$~GeV, where production occurs via freeze-out and $p=-1$ for $T_{\rm RH}\gg m_i \gtrsim  2\times 10^{14}$~GeV, where production occurs via freeze-in (see App.~\ref{app:highTRH_window}). 
Producing the observed baryon asymmetry requires $\epsilon_i\gtrsim 10^{-7},$ which is well within the range shown in Fig.~\ref{fig:eps_contour}.

\section{Proton Decay} \label{sec:Proton_Decay}
Baryon-violation in these models also induces proton decay. At low energies, the relevant operator is
\begin{equation}
    \mathcal{L}_{\slashed{B}} \equiv  {{\cal C}_{iab}}(\bar{u}^a \bar{d}^i) (\bar{u}^b  \bar{e})= \sum_j\frac{(\c_{jia}\kap_b^{j})}{m_{j}^2}(\bar{u}^a \bar{d}^i) (\bar{u}^b  \bar{e}).
\end{equation}
We assume that running the Wilson coefficients from the energy scale of the $\cal S$ particles to $m_p$ contributes at most an $\mathcal{O}(1)$ correction to the derived bounds, which we omit throughout. We also take $\gamma \to 0$ in the discussion for simplicity; $\gamma$ has little effect except for $\gamma \gg y_b^{-2}$ where a cancellation occurs in the proton decay rate as is evident in Fig.~\ref{fig:eps_contour}. The resulting bounds on leptoquark masses and couplings depend on which lepton flavor $\cal S$ couples to. As the lepton flavor structure is free in this construction, $\bar{e}$ could represent electron, muon or tau flavor.

We find that the strongest bounds on couplings to electron flavor come from the proton lifetime constraint $\tau(p\to\bar K^0 e^+)>1.0\times10^{33}\,{\rm yr}$~\cite{Super-Kamiokande:2005lev}. 
Proton decay to the pion and positron is suppressed relative to this by the MFV interactions. The relevant decay width is given by
\begin{equation}
        \Gamma_{p\rightarrow \bar{K}^0 e^+ }~ \approx \frac{m_p}{32\pi}\left(1- \frac{m_{\bar{K}_0}^2}{m_p^2}\right)^2  |W_0|^2 |\mathcal{C}_{211}|^2\,,
\end{equation}
where $W_0\sim 0.1$ GeV$^2$ is the proton decay form factor~\cite{Yoo:2021gql}, and 
\begin{align}
\label{eq:C211}
|\mathcal{C}_{211}| &\approx \frac{4\pi r_\alpha r_\beta}{M^2}\,\frac{y_u y_b}{y_s y_t}\,\frac{|V_{ud}||V_{ub}|}{|V_{us}||V_{td}|}
\approx 9\times10^{-3}\,\frac{r_\alpha r_\beta}{M^2}\,,
\end{align}
 so satisfying the lifetime bound requires
\begin{align}
\label{eq:Mbound}
M\gtrsim 10^{14}\,{\rm GeV}\, (r_\alpha r_\beta)^{1/2}.
\end{align}
Fig.~\ref{fig:eps_contour} shows viable parameter space at low $r_\alpha = r_\beta$, for $M \lesssim 10^{11}$~GeV needed for the low-reheating scenario.

Another avenue is to couple $\S$ only to $\bar \tau$. This renders proton decay harmless even for lighter $\S$. The leading decay channel $p^+ \to \tau^{+*} \to \pi^+\bar{\nu}_\tau$ has width
 \cite{Marciano:1994bg,Heeck:2024jei} 
\be
    \Gamma_{p\rightarrow \pi^+ \bar{\nu}_\tau}  
    \approx \frac{ |V_{ud}|^2 G_F^2 f_\pi^4\,  m_\tau^2}{16\pi m_p  } \frac{(m_p^2-m_\pi^2)^2 }{\left(m_p^2 - m_\tau^2\right)^2 } |W_0|^2 |{{\cal C}_{111}}|^2 ,~~~~~~
\ee
with $f_\pi \approx 130$ MeV. The Wilson coefficient is 
\begin{align}
        |{{\cal C}_{111}}| &\approx  \frac{4\pi r_\alpha r_\beta}{M^2}\,\frac{y_u y_d y_b}{y_s^2 y_t}\,\frac{|V_{ub}|}{|V_{td}|}\approx 10^{-4}\,\frac{r_\alpha r_\beta}{M^2}\,.
\end{align}
The Super-Kamiokande bound,  $\tau(p^+ \to \pi^+ \bar{\nu})> 3.5\times 10^{32}\,{\rm yr}$ \cite{Super-Kamiokande:2025lxa}, then yields the constraint
\begin{align}
M\gtrsim 6\times10^{9}\,{\rm GeV}\, (r_\alpha r_\beta)^{1/2},
\end{align}
so coupling primarily to the $\tau$ considerably weakens the bound from Eq.~\eqref{eq:Mbound}, which assumes a coupling to the electron.

\section{Discussion}
\label{sec:disc}

In this {\it Letter}, we presented a proof-of-principle model of baryogenesis where the SM is extended by a scalar leptoquark with interactions obeying MFV, and where the Jarlskog invariant is the only source of \CP violation. We specified a leptoquark charged as $(\bar{\bm{3}},\bm{1})_{1/3}\times \mathbf{3}_{\bar d}$ under ${\cal G}_{\rm SM}\times{\cal G}_F$. While the same general principles would remain, different charge assignments would alter numerics of the phenomenology significantly and we do not claim a full classification here.

The main challenge facing these models is that renormalizable interactions responsible for baryogenesis violate $B+L$ but preserve $B-L$. Thus, obtaining a baryon asymmetry that survives sphaleron washout requires at least one additional ingredient beyond the scalar leptoquark. One may introduce an additional particle that decays to leptoquarks after sphalerons decouple, or introduce additional lepton violation to washout lepton number from the initial $B+L$ asymmetry, (e.g. by adding right-handed neutrinos to the model). 

Although intended as a proof of principle, this model provides a counterexample to the longstanding expectation that SM \CP violation is insufficient for baryogenesis.
\begin{acknowledgments}
\noindent 
We thank Andrzej Buras, Gino Isidori, Matthew McCullough, Josh Ruderman and Jure Zupan for useful conversations.
This manuscript has been authored in part by Fermi Forward Discovery Group, LLC under Contract No.
89243024CSC000002 with the U.S. Department of Energy, Office of Science, Office of High Energy Physics. HX is supported by the U.S. Department of Energy under grant
DE-SC0026297.  IB is supported by the U.S. Department of Energy under award DE-SC0020262
\end{acknowledgments}

\bibliography{bibl.bib}


\onecolumngrid

\pagebreak
\begin{center}
   \textbf{\large SUPPLEMENTARY MATERIAL \\[.2cm] ``\mytitle''}\\[.2cm]
  \vspace{0.05in}
  {}
\end{center}

\setcounter{equation}{0}
\setcounter{figure}{0}
\setcounter{table}{0}
\setcounter{section}{0}
\setcounter{page}{1}
\makeatletter

\onecolumngrid

{

\renewcommand{\theequation}{A\arabic{equation}}
\setcounter{equation}{0}

\section{Invariants in the Standard Model framework}
\label{app:inv}

In the SM, it is difficult to generate a baryon asymmetry using only the \CP\,violation of the CKM matrix. Since the SM only violates baryon number through  sphaleron-induced processes, an asymmetry can only arise at $T \gtrsim T_{\rm EW} \sim 100$ GeV,  when these reactions are efficient compared to Hubble expansion. However, at these same temperatures, SM dynamics are approximately flavor-universal and all CKM-sensitive processes arise from Yukawa matrix insertions. 
In the fully flavor-blind limit, $Y_{u,d} \to 0$, ${\cal G}_F$ is restored and the CKM matrix becomes unphysical, so no baryon asymmetry can arise.

At high temperatures, the leading CKM-induced \CP\, violation arises from the flavor-invariant~\cite{Jarlskog:1985ht, Jarlskog:1986mm}
\be
\label{eq:XCP}
  X_{\rm CP} = \frac{ v^{12} }{128}  \, \text{Im} \left\{ \text{det} \left[ Y_u Y_u^\dagger, Y_d Y_d^\dagger \right] \right\}  = J \Delta m_{tc}^2  \Delta m_{tu}^2  \Delta m_{cu}^2  \Delta m_{bs}^2  \Delta m_{bd}^2  \Delta m_{sd}^2  ~,
\ee
which depends on both the mass-differences of the SM quarks, $\Delta m_{ij}^2 \equiv  m_i^2-m_j^2 $, and on the Jarlskog rephasing-invariant quantity $J$, where we have followed the conventions of Ref. \cite{Bento:2023owf}. 
Thus, applying dimensional analysis at the weak scale, a naive estimate of the SM baryon asymmetry\footnote{Note that, for this argument, we have also assumed that a sufficient departure from equilibrium occurs at the weak scale, so that all the Sakharov conditions can be satisfied \cite{Sakharov:1967dj}. However, the SM alone does not supply such a departure because the electroweak crossover is not a first-order phase transition \cite{Kajantie:1996mn,DOnofrio:2014rug}, so additional dynamics are required for this as well.} yields $Y_B \sim X_\text{CP}/T_{\rm EW}^{12}
\lesssim 10^{-20},$
which falls far short of the observed value, $Y_B^{\rm obs} \sim 10^{-10}$.

At lower temperatures, SM dynamics are not necessarily flavor-universal and generically depend on multiple energy scales beyond the temperature; combined these features allow \CP-violating observables to evade the suppression in Eq.\eqref{eq:XCP}. For example, in models of Mesogenesis \cite{Elor:2018twp,Nelson:2019fln,Elor:2020tkc,Davoudiasl:2026qba}, the baryon asymmetry arises from particle-antiparticle oscillations, in which the $B$ and $\bar B$ mesons are initially produced as flavor eigenstates with equal number densities. Their subsequent oscillations depend on the CKM phase in addition to particle masses, decay widths, and hadronic matrix elements. In this low-temperature regime, the \CP\, violation from these oscillations cannot be characterized purely as a Yukawa-spurion expansion about a flavor-blind limit in which the temperature is the only dimensionful scale. Nevertheless, in the limit where the CKM phase becomes unphysical (e.g. for degenerate quark masses), the baryon asymmetry still vanishes, but this cancellation occurs without the full product of $\Delta m_{ij}^2$ factors from Eq. ~\eqref{eq:XCP}.

In contrast with both Mesogenesis and the high-temperature SM, the leptoquark model studied here contains additional flavor-charged degrees of freedom ($\mathcal{S}_i$) whose dynamics allow CKM-induced $CP$ violation at high temperatures. Although the ${\cal S}_i$ masses and couplings respect MFV and introduce no additional sources of flavor breaking, they allow new $CP$-violating observables, which are not necessarily proportional to $X_{\rm CP}$. Furthermore, since  $\mathcal{S}_i$ decays  violate baryon number and can occur out-of-equilibrium, the remaining Sakharov conditions can also be satisfied within the same framework.
 Finally, we note that the baryon asymmetry generated in our model vanishes if the ${\cal S}_i$ masses become degenerate, hinting at the existence of a new flavor-invariant quantity in analogy with $X_{\rm CP}$ at scales well above those of the ${\cal S}_i$ masses, $m_i$.  However, exploring the details of such structures is beyond the scope of the present work. 
 
\section{Calculation of Baryon Asymmetry}\label{app:Calculation of Baryon Asymmetry}

Consider decays of a scalar leptoquark $\S_i$ into final states labeled $F$. The decay asymmetry for each component is
\begin{align}
\label{eq:eps_i}
   \epsilon_i = 
   \sum_F B_F \frac{    \Gamma( \S_i \to F ) -  \Gamma(\S_i^\dagger \to \bar F   )}{ \Gamma_{i}}
   =
    \frac{\sum_{F}B_F \left\{|\mathcal{M}( \S_i \rightarrow F )|^2 - |\mathcal{M}( \S_i^\dagger \rightarrow \bar  F )|^2 \right\}}{\sum_{F} |\mathcal{M}( \S_i\rightarrow F )|^2 },
\end{align}
where $B_F$ is the baryon number of decay channel $F$ and $\Gamma_i$ is the total width for $\S_i$. CPT and unitarity force the inclusive rate difference, summed over all final states, to vanish; the baryon-number weighting in Eq.~\eqref{eq:eps_i} is what leaves a nonzero asymmetry. This asymmetry arises from interference between tree and loop-level diagrams in Fig.~\ref{fig:BV_Decay_Diagrams} and we can write the amplitude and its \CP conjugate as 
\begin{align}
\label{eq:amptot}
    \mathcal{M}(\S_i \to F) &=C^i_{T,F}\, \mathcal{A}^i_T + \sum_k \left( 
    C^{ik}_{P,F}\, \mathcal{A}^{ik}_P + C^{ik}_{V,F}\, \mathcal{A}^{ik}_V \right)~ \\ 
      \mathcal{M}(\S_i^\dagger \to \bar F) &=C^{i *}_{T,F}\, \mathcal{A}^i_T + \sum_k \left( 
    C^{ik *}_{P,F}\, \mathcal{A}^{ik}_P + C^{ik *}_{V,F}\, \mathcal{A}^{ik}_V \right)~,
\end{align}
where $k$ indexes contributions from virtual $\S_k$ exchange in loop diagrams and the subscripts $T, P, V$ refer to tree-level, propagator correction, and vertex correction, respectively, as shown in Fig.~\ref{fig:BV_Decay_Diagrams}.\footnote{For a more explicit discussion of \CP-odd and \CP-even phases in interference, see for example Appendix A of~\cite{Bigaran:2024tmp}.} We have factorized each amplitude such that the $C$ coefficients carry all flavor-, color- and other model-dependent contributions, while the $\mathcal{A}$ terms carry the kinematic ones; this factorization is valid only in the limit where SM fermions are treated as massless. To leading order, the numerator of Eq.~\eqref{eq:eps_i} can be written 
\begin{align}
\label{eq:amp-diff}
|\mathcal{M}( \S_i\rightarrow F )|^2 - |\mathcal{M}( \S_i^\dagger \rightarrow \bar{F} )|^2 
    &=  -4 \sum_k \left[  \operatorname{Im}(C^i_{T,F}\, C^{ik *}_{P,F}) \operatorname{Im}(\mathcal{A}^i_T \mathcal{A}^{ik *}_P)  + \operatorname{Im}(C^i_{T,F}\, C^{ik *}_{V,F}) \operatorname{Im}(\mathcal{A}^i_T \mathcal{A}^{ik *}_V)  \right] ,
\end{align}
so combining Eqs. \eqref{eq:eps_i} and \eqref{eq:amp-diff},  
the asymmetry parameter becomes
\begin{align}
    \epsilon_i= -\frac{4\sum_{F}B_F \sum_{L}\sum_k \operatorname{Im}(C^i_{T,F}\, C^{ik *}_{L,F}) \operatorname{Im}(\mathcal{A}^i_T \mathcal{A}^{ik *}_L)}{\sum_{F}|C^i_{T,F}|^2|\mathcal{A}^i_T |^2}, ~~~~
\end{align}
where $L\in\{P,V\}$ runs over propagator and vertex corrections. This recovers Eq.~\eqref{eq:asymm-factor} in the main body. Since this result factorizes kinematics in $\operatorname{Im}(\mathcal{A}_T \mathcal{A}_L^*)$ and MFV couplings in $\operatorname{Im}(C_{T,F}\, C^{*}_{L,F})$, we discuss these terms separately.

\begin{figure}
    \centering
\includegraphics[width=0.75\linewidth]{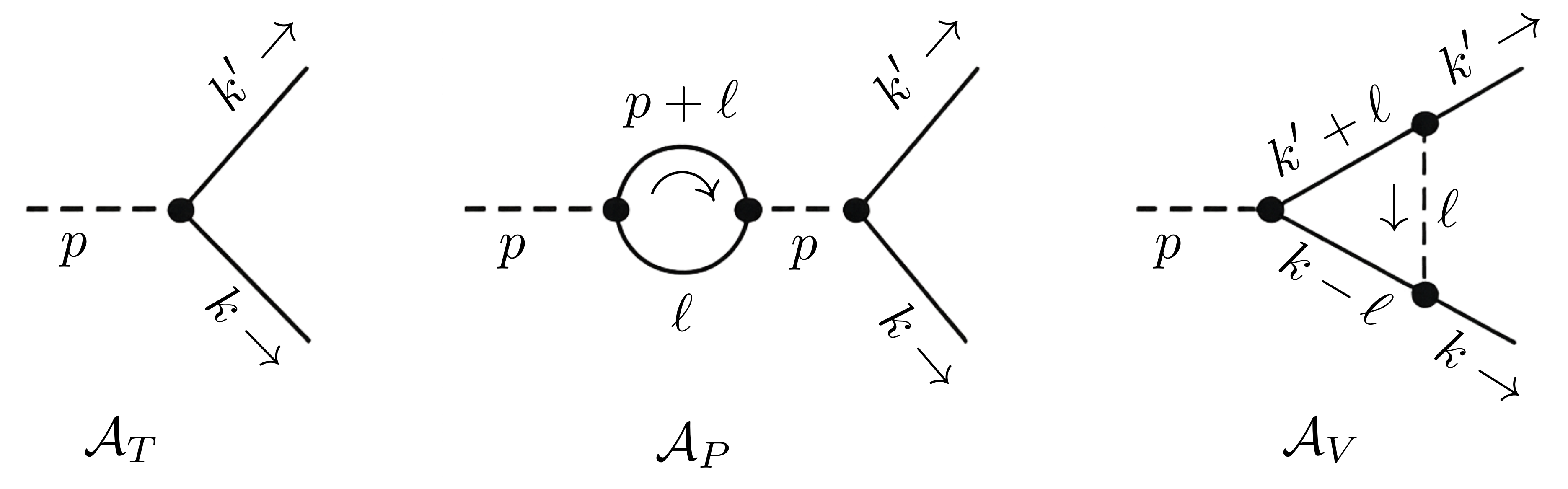}
    \caption{Conventions used to calculate the kinematic parts of Feynman diagrams for $\S_i$ decays, where $p^2 = m_i^2$,  $p = k+k^\prime,  k^2 = k^{\prime 2} = 0$, and scalar internal lines correspond to $\S_k$ propagators with mass $m_k$. Note that for the loop diagrams, we have ${\cal A}_{P,V} \propto {\cal A}_T = \bar u(k) P_R v(k)$ where $u$ and $v$ are Dirac spinors and $T$ and $P,V$ represent tree-level, propagator, and vertex correction terms, respectively. Note that we have suppressed flavor indices here, but the functional form of these expressions applies to every combination we consider here.}
    \label{fig:kinematic_conventions}
\end{figure}
 
\subsection{Kinematic Contributions}
 We follow the subtraction scheme of Ref.~\cite{Bigaran:2024vnl}, defined by requiring that, at zero $p^2$, all loop corrections to the leptoquark vertex are canceled by the counterterm, where $p^2$ is the momentum-squared of the initial-state scalar.
The corrections below are evaluated at the mass scale of the decaying $\S_i$.\footnote{Note that our notation differs from that of~\cite{Bigaran:2024vnl}: in our convention the loop-level contribution does not include the tree-level piece.} For the vertex correction 
\begin{align}
   \frac{ \mathcal{A}^{ik}_V}{\mathcal{A}^{i}_T} =
   i \text{Im}[V_k({m_{i}^2})]
~~~,~~~
    V_k= \frac{1}{16\pi^2}[ b_0 (0; 0, m_k^2)+m_i^2 c_1(m_i^2, 0, 0 ;0, 0, m_k^2)],
\end{align}
 where the two-point function $b_0 (0; 0, m_k^2)$ is explicitly real-valued and does not contribute, and the relevant Passarino-Veltman function is given by
\begin{align}
c_1(m_{i}^2, 0, 0;0, 0, m_{k}^2) = \frac{m_{k}^2}{m_{i}^4} \left\{\text{Li}_2\left(-\frac{m_{i}^2}{m_{k}^2}\right)+ \left(\log\frac{m_{i}^2}{m_{k}^2} -i\pi\right)\left[\log\left(1+\frac{m_{i}^2}{m_{k}^2}\right)-\frac{m_{i}^2}{m_{k}^2}\right]+\frac{m_{i}^2}{m_{k}^2}\right\}, 
\end{align}
so the vertex correction becomes 
\begin{align}
   \frac{ \mathcal{A}^{ik}_V}{\mathcal{A}^{i}_T} = -\frac{i\pi}{16\pi^2} \frac{m_{k}^2}{m_{i}^2} \left[\log\left(1+\frac{m_{i}^2}{m_{k}^2}\right)-\frac{m_{i}^2}{m_{k}^2}\right],
\end{align}
which agrees with the kinematic result of~\cite{Nanopoulos:1979gx} for GUT baryogenesis. 

For the propagator correction, we additionally include the scalar propagator width and loop-induced scalar mixing, which are absent from standard GUT baryogenesis models (e.g.~\cite{Nanopoulos:1979gx}), where such effects could be absorbed into the renormalization scheme. The resulting correction is 

\begin{align}
   \frac{ \mathcal{A}^{ik}_P}{\mathcal{A}^{i}_T}  =
   \frac{i\text{Im}\Sigma(m_i^2)}{m_i^2- m_k^2 +i m_k\Gamma_k}~,
\end{align}
where $\Gamma_k$ is the tree-level width, and
\begin{align}
    \Sigma(p^2)= \frac{p^2}{16\pi^2} \left[ \frac{1}{\epsilon} + 2-\gamma_E + \ln (4\pi)+i\pi -\ln \frac{p^2}{\mu^2}\right],
\end{align}
so that the full expression can be written 
\begin{align} 
\frac{ \mathcal{A}^{ik}_P}{\mathcal{A}^{i}_T} =\frac{m_i^2}{16\pi^2}\frac{i\pi}{m_i^2- m_k^2 +i m_k\Gamma_k}=\frac{m_i^2}{16\pi} \frac{ [i(m_i^2-m_k^2)+m_k \Gamma_k] } {(m_i^2-m_k^2)^2 +m_k^2 \Gamma_k^2}.
\end{align}
This scaling resembles that of resonant leptogenesis models with $\epsilon$-type \CP violation, e.g.~\cite{Pilaftsis:2003gt}, with the difference arising from mixed scalars rather than fermions.

The kinematic contributions are therefore
\begin{align}
    f^{ik}_{P} &\equiv 16\pi\frac{\text{Im}(\mathcal{A}_T^i\mathcal{A}_P^{ik*})}{|\mathcal{A}_T^i|^2} = \frac{ m_i^2(m_k^2-m_i^2) } {(m_k^2-m_i^2)^2 +m_k^2 \Gamma_k^2}\\
    f^{ik}_{V} &\equiv16\pi\frac{\text{Im}(\mathcal{A}_T^i\mathcal{A}_V^{ik*})}{|\mathcal{A}_T^i|^2} = \frac{m_{k}^2}{m_{i}^2} \left[\log\left(1+\frac{m_{i}^2}{m_{k}^2}\right)-\frac{m_{i}^2}{m_{k}^2}\right]
\end{align}
The finite width keeps $f^{ik}_P$ finite near mass degeneracy; at exact degeneracy $m_i=m_k$ it vanishes, and its magnitude is largest just off degeneracy, at $|m_i^2-m_k^2|\sim m_k\Gamma_k$. 

\subsection{Flavor Coefficient Contributions} \label{sec:Flavor Coefficient Contributions}
To calculate the flavor coefficients, we take the leptoquark $\S$ to transform as a $\mathbf{3}_{\bar d}$ under the flavor group ${\cal G}_F$. 
Using this formalism, we express $A_{ija}$ and $B^i_a$ in terms of SM Yukawa couplings and the CKM matrix, and then we relate the \CP asymmetry parameter to the Jarlskog invariant. Throughout this appendix there is \emph{no implied summation over repeated flavor indices}; all flavor summations are written explicitly. We also do not always distinguish superscript from subscript indices.

Explicitly restoring the color structure, the Lagrangian reads
\be  \mathcal{L}_{\mathcal{S}} = \epsilon_{\alpha\beta\gamma}{\c}_{i j a}  \S^{i\alpha}  \bar d^{j\beta} \bar{u}^{a\gamma}+  \delta^\alpha_{\beta}\kap^i_{a} \S^\dagger_{i, \alpha} \bar u^{a\beta} \bar e + h.c.\, ,
\ee
where $\alpha, \beta$ and  $\gamma$ are explicit $SU(3)_c$ color indices. In the main text, the color indices are already fully summed over and included within $|C_{T,F}|^2$ and $\operatorname{Im}(C_{T,F}C_{L,F}^{*})$, so color indices are suppressed from the notation.
Following the index conventions of Fig. \ref{fig:BV_Decay_Diagrams}, the tree-level coefficients in Eq. \eqref{eq:amptot} are 
\be
C^i_{T,u d} = \epsilon_{\alpha \beta\gamma}\sum_{j,a} A_{ija} ~~,~~~
C^i_{T,ue} = \delta^{\alpha}_{\beta}\sum_{a} B^{i*}_{a} ~~,~~~
\ee
The propagator-loop coefficients are 
\be
C^{ik}_{P,ud} &=& \epsilon_{\alpha \beta\gamma} \sum_{j,a,b} B^{i*}_b B^k_{b} A_{kja} 
+ 2 \epsilon_{\alpha \beta\gamma}\sum_{j,a,b,\ell} A_{i\ell b}A^*_{k\ell b} A_{kja}   \\ 
C^{ik}_{P,ue} &=& 2\sum_{j,a,b} A_{ijb} A^*_{kjb} B^{k*}_{a} + \delta^{\alpha}_{\beta}\sum_{a,b} B^{i*}_b B^{k}_b B^{k*}_a
~~,~~~
\ee

which respectively correspond to the middle two diagrams in each row of Fig. \ref{fig:BV_Decay_Diagrams}, and the vertex diagrams are
\be
C^{ik}_{V,ud} = -\epsilon_{\alpha \beta\gamma} \sum_{a,b,j} B^{i*}_b B^k_a A_{kjb} ~~~,~~~~  C^{ik}_{V,ue} = -2 \delta^{\alpha}_{\beta}\sum_{a,b,j}A_{ijb} A^*_{kja} B^{k*}_b~~.
\ee

Note that the subscripts in $ud$ and $ue$ correspond to the processes $\S \to \bar u^\dagger \bar d^\dagger$ and $\S\to \bar u \bar e$ channels, respectively, where bars and daggers are suppressed in the coefficient subscript.

\subsubsection*{\textbf{Coupling Simplification in Down-Aligned Basis}}
The cubic MFV couplings from Eqs.~\eqref{eq:c} and \eqref{eq:kap} can be written 
\begin{equation}\label{eq:couplings}
A_{ija} = \alpha \! \! \sum_{A,B,C}\eps^{ABC}\,(Y_d)_{Ai}\,(Y_d)_{Bj}\,(Y_u)_{Ca}\,,\qquad
B^i_a = \beta\!\!\sum_{A,B,C}\eps_{ABC}\,\sum_{b,c}\eps_{abc}\,(Y_d^\dagger)^{iA} \,(Y_u^\dagger)^{bB}\,(Y_u^\dagger)^{cC}\, ,
\end{equation}
and we work in the \emph{down-aligned basis} in which
\begin{equation}
\label{eq:basis}
Y_d = \hat D = \mathrm{diag}(y_d,\,y_s,\,y_b)\,,\qquad
Y_u = V^\dagger \hat U\,,\qquad
\hat U = \mathrm{diag}(y_u,\,y_c,\,y_t)\,,
\end{equation}
where $V$ is the CKM matrix. We also define the \emph{complement Yukawas} $\ybar_{d_i}$ and $\ybar_{u_a}$ as:
\begin{equation}\label{eq:ybar}
\ybar_{d_i} \equiv \prod_{j\neq i}y_{d_j}\,,\qquad
\ybar_{u_a} \equiv \prod_{b\neq a}y_{u_b}\,.
\end{equation}
For the $A_{ija}$ coupling, the diagonal $Y_d$ sets $A=i$ and $B=j$; the $\eps^{ABC}$ symbol then forces $C$ to be the remaining index $k=\{1,2,3\}\setminus\{i,j\}$. Substituting these relations gives 
\begin{equation}\label{eq:A_closed}
A_{ija}/\alpha \;=\; \sum_k\eps_{ijk}\;\ybar_{d_k}\;y_{u_a}\;V^*_{ak}
\end{equation}
Additionally, in the down-aligned basis we have
\begin{equation}
\label{eq:beta_app}
B^i_a/\beta = y_{d_i}\,\ybar_{u_a}\sum_{B,C}\eps_{iBC}\sum_{b,c}\eps_{abc}\,V_{bB}\,V_{cC}\, ~.
\end{equation}
The $SU(3)$ cofactor identity gives
\begin{align} 
[\mathrm{cof}(V)]_{\alpha\beta} = \tfrac{1}{2}\sum_{\gamma,\delta}\eps_{\alpha\gamma\delta}\sum_{\mu,\nu}\eps_{\beta\mu\nu}\,V_{\gamma\mu}\,V_{\delta\nu}= V_{\alpha \beta}^* ,
\end{align} 
so Eq.~\eqref{eq:beta_app} simplifies to yield  
\begin{equation}\label{eq:B_closed}
\;B^i_a/\beta \;=\; 2\,y_{d_i}\;\ybar_{u_a}\;V^*_{ai}
\end{equation}

\subsubsection*{\textbf{\CP Asymmetry Calculation}}
All \CP violation in this system is proportional to the Jarlskog invariant, which is defined by
\begin{equation}
\label{eq:jarlskog}
\Ima \big(V_{\alpha p}\,V_{\beta q}\,V^*_{\alpha q}\,V^*_{\beta p}\big) \;=\; J \sum_{\gamma, r} \epsilon_{\alpha\beta \gamma} \epsilon_{pqr}
\end{equation}

Here $\gamma$ and $r$ are the indices left free by $\alpha\beta$ and $pq$. Since the loop integral weights each intermediate $k$ differently, we evaluate the vertex, propagator, and tree factors at fixed $k$ before summing over $k$.\\

\begin{itemize}
\item \textbf{Vertex Factors:} Asymmetries from vertex corrections to $\S_i$ decays are proportional to  
\begin{align}\label{eq:Vi}
\operatorname{Im}(C^i_{T,ud}\, C^{ik*}_{V,ud}) &= -\operatorname{Im}(C^i_{T,ue}\, C^{ik*}_{V,ue}) \;=\;  -2\sum_{j,a,b}\Ima\!\big[\,A_{ija}\,(A_{kjb})^*\,(B^{k}_a)^*\,B^i_b\,\big] \\
&= -8|\alpha|^2 |\beta|^2(\det\hat U)^2\sum_{j, m, n}\eps_{ijm}\,\eps_{kjn}\,\ybar_{d_m}\,\ybar_{d_n}\,y_{d_i}\,y_{d_k}\;\Ima\!\bigg(\sum_{a,b}V^*_{am}\,V_{ak}\,V_{bn}\,V^*_{bi}\bigg)\, ,
\end{align}
where $\hat U$ is the diagonal matrix of up-type Yukawa couplings from Eq.~\eqref{eq:basis}, and the  CKM factor can be simplified using unitarity to yield  
\begin{equation}\label{eq:unitarity_kill}
\sum_{a,b}V^*_{am}\,V_{ak}\,V_{bn}\,V^*_{bi}
\;=\;\bigg(\sum_a V^*_{am}\,V_{ak}\bigg)
\;\bigg(\sum_b V_{bn}\,V^*_{bi}\bigg)\;=\;\delta_{mk}\,\delta_{ni}\; ,
\end{equation}
which has no imaginary part, so the vertex contribution vanishes for all~$i,\,k$.\\

\item {\textbf{Propagator Factors:}}  Asymmetries from propagator corrections to $\S_i$ decays coming with either four $A$ coefficients or four $B$ coefficients contribute no imaginary part as they are proportional to ${\rm Im}(|\sum_{\ell,b} A_{i\ell b}A^*_{k\ell b}|^2)$ and ${\rm Im}(|\sum_{b} B^{i*}_{b}B^{k}_{b}|^2)$ respectively. Non-vanishing contributions are proportional to 

\begin{align}\label{eq:Pi_general}
\operatorname{Im}(C^i_{T,ud}\, C^{ik*}_{P,ud})  &= -\operatorname{Im}(C^i_{T,ue}\, C^{ik*}_{P,ue})  \;=\; 2\sum_{j,a,b}\Ima\!\big[\,B^i_a\,(B^k_a)^*\,A^*_{kjb}\,A_{ijb}\,\big] \notag\\
&=8|\alpha|^2 |\beta|^2\sum_{j, m, n} \varepsilon_{ijm}\varepsilon_{kjn} y_{d_i} y_{d_k} \bar y_{d_m} \bar y_{d_n}\sum_{a,b} \bar y_{u_a}^2 y_{u_b}^2 \Ima\!\bigg(V^*_{ai}\,V_{ak}\,V_{bn}\,V^*_{bm}\bigg)\, \notag \\
&=8|\alpha|^2 |\beta|^2\sum_{j, m, n} \varepsilon_{ijm}\varepsilon_{kjn} y_{d_i}^2   y_{d_j}^2 y_{d_k}^2 \sum_{a,b} \bar y_{u_a}^2 y_{u_b}^2 \Ima\!\bigg(V^*_{ai}\,V_{ak}\,V_{bn}\,V^*_{bm}\bigg)\,
\end{align}
For $i=k$, $m = n$ and the CKM combination is real (see Eq.~\eqref{eq:unitarity_kill}). For $i\neq k$, $m=k$ and $n = i$, giving  
\begin{align}\label{eq:Pi_reduced}
\operatorname{Im}(C^i_{T,ud}\, C^{ik*}_{P,ud}) 
 &\; = -8|\alpha|^2 |\beta|^2({\rm det} \hat{D})^2\sum_{a,b} \bar y_{u_a}^2 y_{u_b}^2 \underbrace{\Ima\!\bigg(V^*_{ai}\,V_{ak}\,V_{bi}\,V^*_{bk}\bigg)}_{=-J\sum_{l, r} \varepsilon_{abl}\varepsilon_{ikr}}\\
&= -8|\alpha|^2 |\beta|^2({\rm det} \hat{D})^2 (y_t^2-y_c^2)(y_t^2-y_u^2)(y_c^2-y_u^2) J \sum_{r}\varepsilon_{ikr}\\
&\approx -8|\alpha|^2 |\beta|^2y_d^2y_s^2y_b^2 y_t^4y_c^2 J \sum_{ r}\varepsilon_{ikr},
\end{align}
where we have used the 
diagonal down-type Yukawa matrix $\hat D$ from Eq.~\eqref{eq:basis}.

\item {\textbf{Tree-Level Coefficients:}} At tree level, the squared amplitude coefficients are  
\begin{align}
    &|C^i_{T,ud}|^2 =2 \sum_{j,a} |A_{ija}|^2 = 2|\alpha|^2 \sum_{m\neq i}\sum_a\bar y_{d_m}^2y_{u_a}^2 |V_{am}|^2\\
    &|C^i_{T,ue}|^2 = \sum_{a} |B^i_{a}|^2 = 4|\beta|^2y_{d_i}^2\sum_{a} \bar y_{u_a}^2 |V_{ai}|^2~,
\end{align}
which agree with the expressions  Eqs.~\eqref{eq:Cud} and \eqref{eq:Cue} of the main text. 
\end{itemize}

\subsection{Final Result} \label{sec:Final Result}
 
The average baryon asymmetry per decay $\epsilon_i$ combines the kinematic factors with the flavor coefficients. The vertex term vanishes by Eq.~\eqref{eq:unitarity_kill}, leaving only the propagator correction. Summing over intermediate states $k \neq i$ gives
\begin{align}
    \epsilon_i  &=
-\frac{4\sum_{F}B_F \sum_{L}\sum_k \operatorname{Im}(C^i_{T,F}\, C^{ik *}_{L,F}) \operatorname{Im}(\mathcal{A}^i_T \mathcal{A}^{ik *}_L)}{\sum_{F}|C^i_{T,F}|^2|\mathcal{A}^i_T |^2},\\
    &= -\frac{(B_{\bar u^\dagger \bar d^\dagger} - B_{\bar u \bar e} )}{4\pi}   
    \frac{\sum_{ k \neq i}\operatorname{Im}(C^i_{T,ud}\, C^{ik*}_{P,ud}) f_{P}^{ik}  }{|C^i_{T,ud}|^2 + |C^i_{T,ue}|^2 }, \\
    &= \frac{2|\alpha|^2 |\beta|^2y_d^2\,y_s^2\,y_b^2\,y_c^2\,  y_t^4\, J}{\pi}   
    \frac{\sum_{r, k \neq i}\varepsilon_{ikr}f_{P}^{ik}  }{ |C^i_{T,ud}|^2 + |C^i_{T,ue}|^2}.
\end{align}

\section{High-Reheat Conversion and Timing Estimates}\label{app:highTRH_window}

\noindent\emph{\textbf{Overview:}} This appendix details the high-reheat scenario of Sec.~\ref{sec:highTRH}, largely reviewing and re-framing the argument in Ref. ~\cite{Fukugita:2002hu}. The approximate treatment below is meant to demonstrate a viable path towards generating the observed baryon asymmetry with a high reheat temperature and an equilibrium initial condition for the ${\cal S}_i$ leptoquark fields.

The scenario proceeds through several events, ordered by the cosmological temperature at which each occurs:
\begin{enumerate}
    \item The universe is initially reheated to a temperature of $T_{\rm RH} > m_i$ and the ${\cal S}_i$ are in equilibrium with the SM plasma as relativistic species.
    
    \item As the universe cools, the ${\cal S}_i$ freeze out at $T\sim m_i/10$ and subsequently decay out of equilibrium at $T_{\cal S} > 10^{12}$ GeV to generate a $B+L$ asymmetry.

    \item Lepton-violating (but baryon-preserving) interactions are efficient in between $ T_{\cal S} >T > 10^{12}$ GeV and converts some of the pure $B+L$ asymmetry into a $B-L$ asymmetry by washing out lepton number. These interactions freeze out at $T^{\rm dec}_{\Delta L =2}$, which occurs before $T\sim 10^{12}$ GeV.

    \item When sphalerons become active at $T_{\rm sph} \sim 10^{12}$ GeV, the remaining $B+L$ is erased and the but the $B-L$ asymmetry of the universe is preserved for all later times.
\end{enumerate}
Schematically, the hierarchy of temperatures is 
\begin{align}
    T_{\rm RH}\gtrsim T_\S\gtrsim T_{\Delta L=2}^{\rm dec} \gtrsim T_{\rm sph},
\end{align}
where the last three of these are determined by setting the corresponding reaction rates equal to the rate of Hubble expansion, $\Gamma(T) = H(T)$, with
$H^2(T)=g_*\pi^2 T^4/90M_P^2$; $g_*=106.75$ is used for our numerical estimates. We justify this event order below after sketching the asymmetry origin.

Let $Y_{\S_i}\equiv n_{\S_i}/s$ denote the abundance just before each $\S_i$ decays. Since these
decays conserve $B-L$, they source a net $B+L$ yield, which satisfies
\begin{align}
Y_{B+L}  &\equiv 2\sum_i Y_{\S_i}\,\epsilon_i,\qquad
Y _{B-L}=0 ,
\label{eq:highTRH_source}
\end{align}
where the factor of two in the left expression arises because $L=B$, which follows from $B-L = 0$, and $\epsilon_i$ is given in Eq.~\eqref{eq:eps_i}. In the absence of additional dynamics at these high temperatures, once sphalerons come into equilibrium, they would erase the pure $B+L$ asymmetry generated from ${\cal S}$ decays. In order to preserve a baryon asymmetry down to lower temperatures, at least some fraction of this $B+L$ asymmetry must be converted into  a $B-L$ asymmetry which sphalerons cannot erase. 

Fortunately, there is a simple mechanism that can convert $B+L \to B-L$: additional right-handed neutrinos with lepton-violating Majorana masses can source $\Delta L=2$ interactions, which can efficiently wash out lepton number before sphalerons equilibrate. Then, after sphalerons equilibrate we have
\begin{align}
Y_B=\frac{28}{79}\kappa (1-r_L)\sum_i Y_{\S_i}\,\epsilon_i .
\label{eq:highTRH_conversion}
\end{align}
where $28/79$ is the sphaleron conversion of $B-L$ into $B$~\cite{Weinberg:2008zzc} and $r_L$ is the surviving fraction of the source lepton number, and $\kappa$ is the fraction of the source protected as $B-L$. 

As we show below, every ingredient that appears in Eq.~\eqref{eq:highTRH_conversion} can be sufficiently large  to yield the observed baryon asymmetry at late times. Specifically, using approximate estimates for the parameters $\kappa$, $r_L$, and $Y_{\mathcal{S}_i}$, the observed baryon asymmetry is achievable for $\epsilon_i\gtrsim 10^{-7}$, which can readily be accommodated within the leptoquark model discussed in the main text.

\medskip
\noindent\emph{\textbf{Chemical Equilibrium Factor ($\kappa$): }} Near equilibrium, the asymmetries are small, so the chemical potentials $\mu_a$ (with $n_a = g_a \mu_a$) are determined by a finite system of equations governed by the fastest interactions. At $T \sim T_\S \sim 10^{14}$~GeV, the only fast processes besides gauge interactions are $\Delta L = 2$ and top-Yukawa mediated processes, i.e. 
\begin{align}
    &\mu_{Q_3} - \mu_{\bar{t}} - \mu_H = 0\\
    &\mu_{L_\alpha} + \mu_H = 0\quad ({\rm for~} \alpha \leq {\rm rank}(m_\nu))
\end{align}
with $(g_{Q_3},g_{\bar{t}},g_{L_\alpha},g_H)=(6,3,2,4)$. The decays deposit baryon number into the right-handed quarks (predominantly $\bar{t}$) and to right handed charged-leptons $\bar{e}$ such that $B-L = 0$. Hypercharge neutrality also, fixes 
\begin{align}
    \sum_a g_a {\cal Y}_a \mu_a = 0\,,
\end{align}
where ${\cal Y}_a$ is the hypercharge. Solving the system of equations gives
\begin{align}
    \kappa \equiv \frac{Y_{B-L}}{\sum_i Y_{\S_i} \epsilon_i } = \frac{2\,{\rm rank}(m_\nu)}{3({\rm rank}(m_\nu)+1)} = \left[ \frac{4}{9},\,\frac{1}{2} \right], \quad \text{for } \mathrm{rank}(m_\nu) = 2,\,3\text{ respectively.}
\end{align}

\medskip
\noindent\emph{\textbf{Lepton Survival Factor ($r_L$): }} If the right-handed neutrinos are heavy compared to the temperature,\footnote{This assumption is made for simplicity and can be relaxed.} the effective lepton-violating interaction can be written 
\be
\mathcal L_{\Delta L=2}=\frac{1}{v^2} m_\nu^{\alpha \beta}(\ell_\alpha H)(\ell_\beta H),
\ee
which governs $\Delta L = 2$ processes. For this choice of interaction, lepton-violating interactions and their decoupling temperature satisfy
\begin{align}
\Gamma_{\Delta L=2}(T)
&\approx \frac{0.12}{4\pi}\frac{\bar m_\nu^2}{v^4}T^3\implies 
T_{\Delta L=2}^{\rm dec}\approx
5.4\times10^{13}\,{\rm GeV}
\left(\frac{{0.1\,\rm eV}}{\bar m_\nu}\right)^2,
\label{eq:DeltaL2_rate}
\end{align}
where $\bar{m}_\nu^2 \equiv \Tr(m_\nu^\dagger m_\nu)$, and for neutrino masses that satisfy laboratory bounds, we demand $T_{\Delta L=2}^{\rm dec}> T_{\rm sph}$.\footnote{Note that if right handed neutrinos are lighter, this need not always occur.} With these assumptions and model ingredients, the lepton-number survival factor is 
\begin{align}
r_L\approx
\exp\left[-\frac{{\rm min}\left(T_\S ,\, m_N\right)}{T_{\Delta L=2}^{\rm dec}}\right]\,,
\label{eq:DeltaL2_efficiency}
\end{align}
and for $m_N\gtrsim T_\S$, efficient washout requires $T_\S \gtrsim T_{\Delta L=2}^{\rm dec}$.

\bigskip
\noindent \emph{\textbf{B+L Washout Restriction: }}Consider $\S_1$ decays for concreteness. We define
\begin{align}
    K_1 = \frac{\Gamma_1}{H(T=m_1)} \approx (26 r_\alpha^2 + 9 r_\beta^2)\left(\frac{m_1}{10^{15}\, \rm GeV}\right)^{-1}
\end{align}
The decay temperature is then $T_{\S_1} = m_1 \sqrt{K_1}$. Note that washout of the $B+L$ asymmetry from inverse decay processes involving $\S_1$ can only be ignored if $K_1\lesssim 1$, but we just noted that we also require $T_\S\gtrsim T_{\Delta L=2}^{\rm dec}$. These conditions are met, for example, with $m_1 \sim  3\times 10^{14}$~GeV and $r_\alpha = r_\beta \approx 0.1$. In general, neither of these two ``bounds'' are hard cutoffs and failing either bound simply requires a more detailed calculation of washout and using $(1-r_L) \approx {\rm min}\left(T_\S ,\, m_N\right)/T_{\Delta L=2}^{\rm dec}$; neither or these complications is fatal to the mechanism.

\medskip
\noindent \emph{\textbf{Initial} $\mathbf{\cal{S}}$ \textbf{Abundance: }}Finally, the leptoquark yield at the time of decay, $Y_{\S_i}$, follows from a standard freeze-out calculation. Note that at these high temperatures, even strong interactions struggle to keep non-relativistic particles in equilibrium since the Hubble rate is extremely high, so $\S_i$ decouples at $T_{\rm FO,\,i}\sim m_i/\mathcal{O}(\rm few)$ with 
\begin{align}
    Y_{\S_i} \sim \frac{H}{s\expval{\sigma v}}\Big|_{T\sim m_i} \sim \frac{m_i}{\alpha_s^2M_P \sqrt{g_*}}\sim 10^{-2} \left(\frac{m_i}{ 2\times 10^{14}~{\rm GeV}}\right)\qquad {\rm for}~m_i \lesssim 2\times 10^{14}~{\rm GeV}~,
    \label{eq:YS_freezeout}
\end{align}
which can therefore be large; however, it can never exceed the relativistic freeze-out yield $Y_{\rm rel}= 45g_{\cal S}\zeta(3)/2\pi^4 g_{*s}\approx 0.01$ which is saturated at the benchmark mass. This benchmark value does not impose an upper bound on the asymmetry generated within this model, but only on the validity of Eq.~\eqref{eq:YS_freezeout} as it governs the mass at which  ${\cal S}$ annihilation at $T\sim m_i$ become slower than Hubble. A heavier $\S_i$ never reaches chemical equilibrium, and its abundance is populated through sub-Hubble freeze-in reactions, which satisfy
\begin{align}
    Y_{\S_i} \sim Y_{\rm rel}\,\frac{\Gamma_{\rm ann}}{H}\Big|_{T\sim m_i} \sim 10^{-2} \left(\frac{2\times 10^{14}~{\rm GeV}}{m_i}\right)\qquad {\rm for}~2\times 10^{14}~{\rm GeV}\lesssim m_i \ll T_{\rm RH}~.
\end{align}

\end{document}